\documentclass[10pt,conference]{IEEEtran}

\usepackage{verbatim}
\usepackage{tabularx}
\usepackage{soul}
\usepackage{enumitem}
\usepackage{booktabs}
\usepackage{blindtext}
\usepackage{multirow}
\usepackage{listings}
\usepackage{caption}
\usepackage{subcaption}
\usepackage{url}
\usepackage{graphicx}
\usepackage[dvipsnames]{xcolor}
\usepackage{makecell}
\usepackage{svg}
\usepackage[most]{tcolorbox}
\usepackage{tikz}
\usepackage{hyperref}
\usepackage{cleveref}

\definecolor{darkgreen}{rgb}{0.0, 0.5, 0.0}

\lstdefinestyle{JavaStyle}{
  language=Java,
  frame=single,
  backgroundcolor=\color{white},
  basicstyle=\footnotesize\ttfamily,
  keywordstyle=\color{blue}\bfseries,
  commentstyle=\color{darkgreen},
  stringstyle=\color{red},
  numbers=left,
  numberstyle=\tiny\color{gray},
  stepnumber=1,
  numbersep=5pt,
  showspaces=false,
  showstringspaces=false,
  showtabs=false,
  tabsize=2,
  captionpos=b,
  breaklines=true,
  breakatwhitespace=true,
  escapeinside={\%*}{*)},
  morekeywords={*,...}
}
\definecolor{mycolor}{RGB}{119,195,236}

\newtcolorbox{highlight}[1][]{
  colback=mycolor!10!white,
  colframe=mycolor!80!black,
  fonttitle=\bfseries,
  coltitle=mycolor!80!black,
  colbacktitle=mycolor!20!white,
  boxrule=1pt,
  arc=3pt,
  outer arc=3pt,
  boxsep=0pt,
  left=10pt,
  right=10pt,
  top=6pt,
  bottom=6pt,
  toptitle=2pt,
  bottomtitle=2pt,
  lefttitle=2pt,
  righttitle=2pt,
  titlerule=0pt,
  attach boxed title to top left={yshift=-\tcboxedtitleheight/2, xshift=2mm},
  boxed title style={arc=3pt, outer arc=3pt},
}

\usepackage{soul}

\graphicspath{ {figures/} }

\def\BibTeX{{\rm B\kern-.05em{\sc i\kern-.025em b}\kern-.08emT\kern-.1667em\lower.7ex\hbox{E}\kern-.125emX}}

\newcommand{\blue}[1]{\colorbox{CornflowerBlue!50}{#1}}
\newcommand{\green}[1]{\colorbox{SpringGreen!50}{#1}}
\newcommand{\yellow}[1]{\colorbox{Goldenrod!50}{#1}}
\newcommand{\red}[1]{\colorbox{Salmon!50}{#1}}
\newcommand{\purple}[1]{\colorbox{Purple!50}{#1}}

\newcommand{\change}[1]{{\color{black} #1}}

\newcommand{\tool}{\textsc{CAT-LM}\xspace}

\newcommand{\cat}{\includegraphics[scale=0.16]{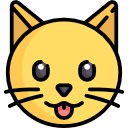}}

\begin{document}
%\fancyhead{}

%
% The "title" command has an optional parameter, allowing the author to define a "short title" to be used in page headers.

%\title{\textsc{CAT-LM}\emoji{smiley-cat}  Training Language Models on Code Aligned with Tests}
\title{\textsc{CAT-LM}\cat \hspace{0.01cm} Training Language Models on Aligned Code And Tests}

\author{
\IEEEauthorblockN{Nikitha Rao\IEEEauthorrefmark{1}, Kush Jain\IEEEauthorrefmark{1}, Uri Alon, Claire Le Goues, Vincent J. Hellendoorn}
\IEEEauthorblockA{\textit{Carnegie Mellon University} \\ {{United States}} \\
\{nikitharao, kdjain, urialon, legoues, vhellendoorn\}@cmu.edu}
}

%
% The "author" command and its associated commands are used to define the authors and their affiliations.
% Of note is the shared affiliation of the first two authors, and the "authornote" and "authornotemark" commands
% used to denote shared contribution to the research.

%
% By default, the full list of authors will be used in the page headers. Often, this list is too long, and will overlap
% other information printed in the page headers. This command allows the author to define a more concise list
% of authors' names for this purpose.

\maketitle

\begingroup\renewcommand\thefootnote{\IEEEauthorrefmark{1}}
\footnotetext{Equal contribution}

% The abstract is a short summary of the work to be presented in the article.
\begin{abstract}
Testing is an integral but often neglected part of the software development process.
Classical test generation tools such as EvoSuite generate behavioral test suites by optimizing for coverage, but tend to produce tests that are hard to understand.
Language models trained on code can generate code that is highly similar to that written by humans, but current models are trained to generate each file separately, as is standard practice in natural language processing, and thus fail to consider the code-under-test context when producing a test file.
In this work, we propose the Aligned \underline{C}ode \underline{A}nd \underline{T}ests Language Model (\tool), a GPT-style language model with 2.7 Billion parameters, trained on a corpus of Python and Java projects. We utilize a novel pretraining signal that explicitly considers the mapping between code and test files when available. We also drastically increase the maximum sequence length of inputs to 8,192 tokens, 4x more than typical code generation models, to ensure that the code context is available to the model when generating test code.
We analyze its usefulness for realistic applications, showing that sampling with filtering (e.g., by compilability, coverage) allows it to efficiently produce tests that achieve coverage similar to ones written by developers while resembling their writing style.
By utilizing the code context, \tool generates more valid tests than even much larger language models trained with more data (CodeGen 16B and StarCoder) and substantially outperforms a recent test-specific model (TeCo) at test completion.
Overall, our work highlights the importance of incorporating software-specific insights when training language models for code and paves the way to more powerful automated test generation.
\end{abstract}


%
% This command processes the author and affiliation and title information and builds
% the first part of the formatted document.
 \begin{figure*}
    %  [trim={left bottom right top},clip]
    \center
    \includegraphics[width=\textwidth,trim={1.5cm 4cm 1.5cm 3.5cm}, clip]{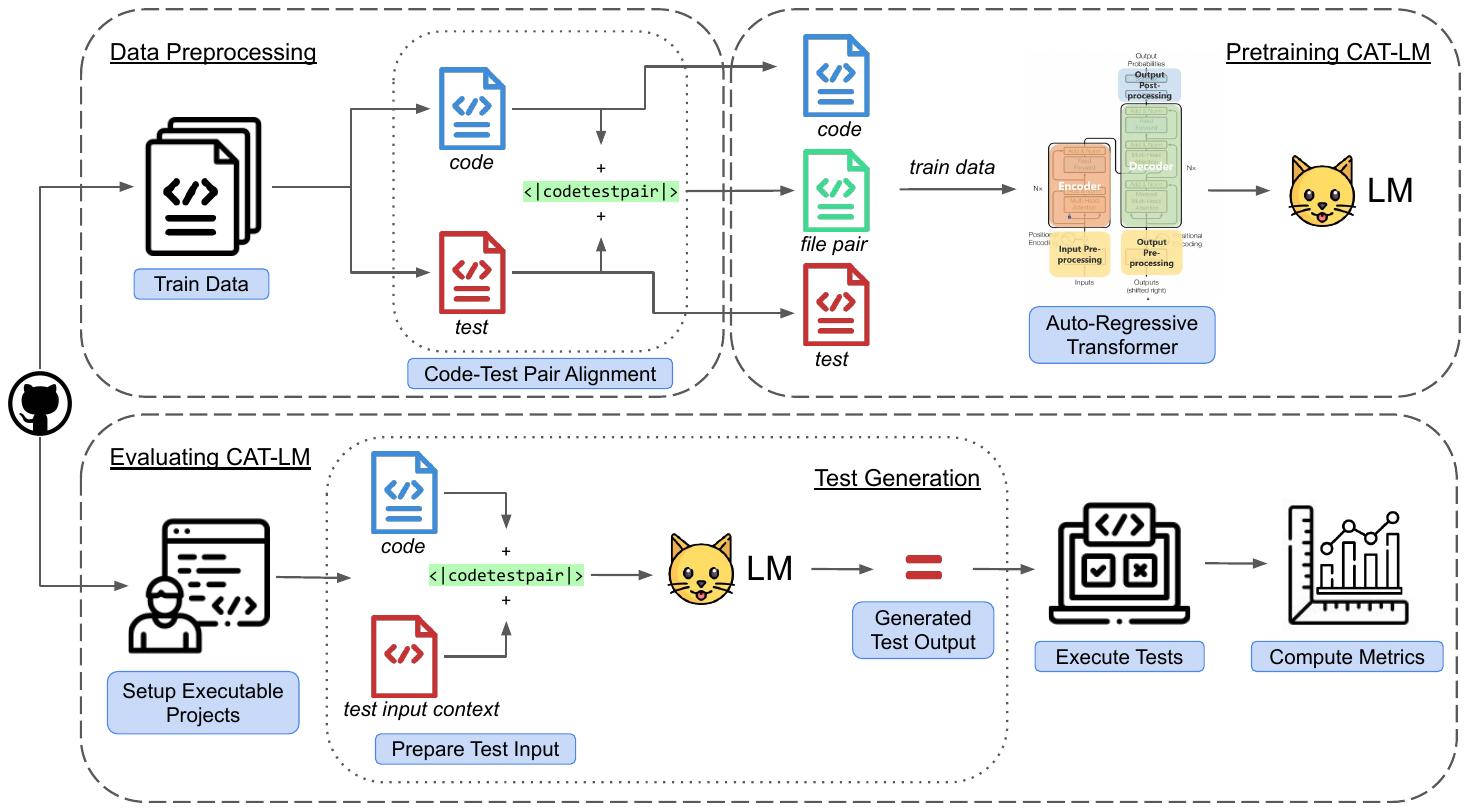}
    \caption{Approach overview. We extract Java and Python projects with tests from GitHub and heuristically align code and test files (top), which, along with unaligned files, train \tool, a large, auto-regressive language model. We evaluate \tool's generated tests on a suite of executable projects (bottom), measuring its ability to generate syntactically valid tests that yield coverage comparable to those written by developers.}
    \label{fig:overview}
  \end{figure*}
  
\section{Introduction}

Software testing is a critical component of the software development process. In well-tested projects, most code files are paired with at least one corresponding test file that implements unit and/or integration test functions that evaluate the functionality of the code. However, writing high quality tests can be time-consuming ~\cite{Beller1, Beller2} and is often either partially or entirely neglected. This has led to extensive work in automated test generation, including both classical~\cite{DinellaTOGA, FraserEvoSuite, BrandtEvoSuiteStudy, BaldoniSymbolicExecution} and neural-based methods ~\cite{DinellaTOGA, WatsonATLAS, VillmowContest}. 

Classical test generation tools like EvoSuite~\cite{FraserEvoSuite} directly optimize to generate high-coverage tests. However, the generated tests are often hard to read and may be unrealistic or even wrong \cite{panichella2020revisiting}. This requires time and effort from developers to verify generated test correctness~\cite{BrandtEvoSuiteStudy}.

Meanwhile, Large Language Models (LLMs) trained on code have made major strides in generating human-like, high-quality functions based on their file-level context ~\cite{Chen2021Codex, CodeGen, FriedInCoder, FiMOpenAI}. Tools like Copilot excel at code generation, and can significantly improve the productivity of its users~\cite{copilot}. Currently, these models are less well-suited for test generation, because they tend to be trained to generate the code in each file separately, standard practice in natural language processing.
Generating meaningful tests, of course, critically requires considering the alignment between the tests and the corresponding code under test. 
Some prior work on neural-based test generation methods has focused on modeling this alignment~\cite{DinellaTOGA, WatsonATLAS, NieETAL22Teco}. However, this work typically focuses on the relatively narrow task of generating individual assertions in otherwise complete tests, based on a single method under test.
Unlocking the more impactful ability to generate entire tests requires leveraging both the entire code file and existing tests as context, which in turn requires substantially larger models.

In this work, we make a significant step towards accurate whole-test generation via \tool, a language model trained on aligned \textbf{C}ode \textbf{A}nd \textbf{T}ests. \tool is a bi-lingual GPT-style LLM with 2.7B parameters. It is trained on a large corpus of Python and Java projects using a novel pretraining signal that explicitly considers the mapping between code and test files, when available, while also leveraging the (much larger) volume of untested code.
Modeling the code file along with the test leads to additional challenges regarding a model's context length. Most code generation models support a context window of up to 2,048 tokens. However, our data analysis indicates that many code-test file \emph{pairs} comprise more than 8K tokens. We thus increase the maximum sequence input length, training \tool with a context window of 8,192 tokens.

Our results show that the model effectively leverages the code file context to generate more syntactically valid tests that achieve higher coverage.
The model provides a strong prior for generating plausible tests: combined with basic filters for compilability and coverage, \tool frequently generates tests with coverage close to those written by human developers.

We evaluate \tool against several strong baselines across two realistic applications: test method generation and test method completion. For test method generation, we compare \tool to both human written tests as well as the tests generated by StarCoder~\cite{starcoder} and, the CodeGen~\cite{CodeGen} model family, which includes mono-lingual models trained on a much larger budget than ours. We also compare against TeCo~\cite{NieETAL22Teco}, a recent test-specific model, for test completion. 
\tool generates more valid tests on average than StarCoder and all CodeGen models, and substantially outperforms TeCo at test completion. Our results highlight the merit of combining the power of large neural methods with a pretraining signal based on software engineering expertise---in this case, the importance of the relation between code and test files.

In summary, we make the following contributions:
\begin{itemize}
    \item We release a corpus of 1.1M code-test file pairs along with 14.4M Java and Python files across 196K open-source projects. We believe this corpus will be useful for many testing-related tasks. 
    \item We release \tool, the first pretrained LLM that models aligned code and test files from Java and Python projects on GitHub.
    \item We release the testing framework used to evaluate the tests generated by \tool.
    \item We conduct an extensive evaluation of \tool with strong baselines on downstream tasks such as test method generation and test completion. 
\end{itemize}

The model checkpoints along with usage code samples can be found at \url{https://github.com/RaoNikitha/CAT-LM}.

\section{Overview}

\tool is a GPT-style model that can generate tests given code context. \Cref{fig:overview} shows an overview of our entire system, which includes data collection and preprocessing (detailed in \Cref{subsec:data}), pretraining \tool (\Cref{sec:training}), and evaluation (\Cref{sec:evaluation}). 

We first collect a corpus of ca. 200K Python and Java GitHub repositories, focusing on those with at least 10 stars. We split these at the project level into a train and test set (\Cref{subsec:data}). We filter our training set following CodeParrot~\cite{CodeParrot} standards (including deduplication), resulting in $\sim$15M code and test files. We align code and test files using a fuzzy string match heuristic (\Cref{subsec:preprocessing}).

We then prepare the training data, comprising of the code-test file pairs, paired with a unique token (\texttt{<|codetestpair|>}), as well as unpaired code and test files. We tokenize the files using a custom-trained sentencepiece tokenizer~\cite{spm}. We then determine the appropriate model size, 2.7B parameters based on our training budget and the Chinchilla scaling laws \cite{chinchilla}. We use the GPT-NeoX toolkit~\cite{neox} enhanced with Flash Attention~\cite{dao2022flashattention} to pretrain \tool using an auto-regressive (standard left-to-right) pretraining objective that captures the mapping between code and test files, while learning general code and test structure. 

Finally, we evaluate \tool on the held-out test data. We manually set up all projects with executable test suites from the test set to form our testing framework. We prepare our test inputs for \tool by concatenating the code context to the respective test context for test generation. The test context varies based on the task. We asses our model's ability to generate (1) the first test method, (2) the last test method, add (3) an additional, new test to an already complete test suite. We also evaluate completing a statement within a test function. We tokenize prepared input and task \tool with sampling multiple (typically 10) test outputs, each consisting of a single method. We then attempt to execute the generated tests with our testing framework and compute metrics like number of generated tests that compile and pass, along with the coverage they provide, to evaluate test quality.
  
\section{Tasks}
\label{sec:tasks}

 \begin{figure}
    %  [trim={left bottom right top},clip]
    \center
    \includegraphics[width=0.35\textwidth, trim={1.5cm 7cm 19cm 4cm},clip]{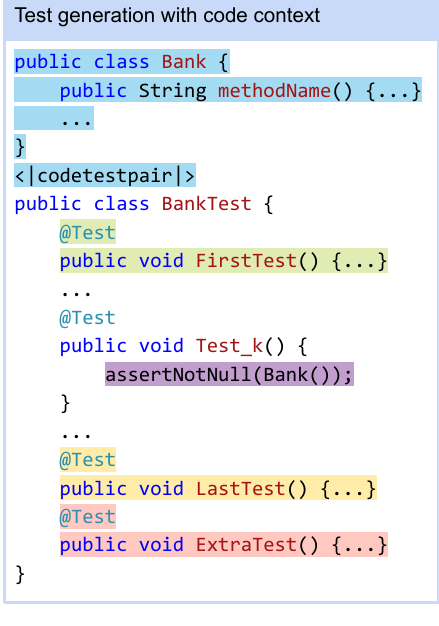}
    \caption{Evaluation tasks, with \blue{code context} shown for completeness: test generation for the \green{first test method}, \yellow{last test method}, and \red{extra test method}, along with \purple{test completion} for Java.}
    \label{fig:tasks}
  \end{figure}

We describe two tasks for which \tool can be used, namely test method generation (with three settings) and test completion. \Cref{fig:tasks} demonstrates the setup for all tasks including code context.

\subsection{Test Method Generation}
\label{sec:testmethodgen}

Given a partially complete test file and its corresponding code file, the goal of \emph{test method generation} is to generate the next test method. Developers can use test generation to produce an entire test suite, or add tests to an existing test suite to test new functionality. We evaluate three different settings, corresponding to different phases in the testing process, namely generating (1) the \emph{first test} in the file, representing the beginning of a developer's testing efforts. In this setting, we assume that basic imports and high-level scaffolding are in place, but no test cases have been written, (2) the \emph{final test} in a file, assessing a model's ability to infer what is missing from a near-complete test suite. We evaluate this ability only on test files that have two or more (human-written) tests to avoid cases where only a single test is appropriate, and (3) an \emph{extra} or additional test, which investigates whether a model can generate new tests for a largely complete test suite. Note that this may often be unnecessary in practice.

\subsection{Test Completion}
\label{sec:testcompletion}

The goal of \emph{test completion} is to generate the next statement in a given incomplete test method. Test completion aims to help developers write tests more quickly.
Although test completion shares similarities with general code completion, it differs in two ways: (1) the method under test offers more context about what is being tested, and (2) source code and test code often have distinct programming styles, with test code typically comprising setup, invocation of the method under test, and assertions about the output (the test oracle).
{\renewcommand{\arraystretch}{1.2}
\begin{table}[t]
{\small
    \begin{center}
    \centering
    \caption{Summary statistics of the overall dataset.}
    \label{tab:data}
    \begin{tabular}{@{}l@{\hspace{1mm}}|@{\hspace{1mm}}l@{\hspace{1mm}}|@{\hspace{2mm}}r@{\hspace{1mm}}|@{\hspace{2mm}}r@{\hspace{1mm}}|@{\hspace{2mm}}r@{\hspace{1mm}}}
        \toprule

         \multicolumn{2}{@{\hspace{1mm}}c@{\hspace{1mm}}|@{\hspace{2mm}}}{\textbf{Attribute}} & \textbf{Python} & \textbf{Java} & \textbf{Total} \\
        \midrule
        \multirow{4}{*}{Project} & Total & 148,605 & 49,125 & 197,730  \\
        & Deduplicated & 147,970 & 48,882 & 196,852\\
        & W/o Tests & 84,186 & 15,128 & 99,314\\
        & W/o File pairs & 108,042 & 23,933 & 131,975 \\
        \hline
        \multirow{2}{*}{\makecell{Size\\(GB)}} & Raw & 123 & 157 & 280 \\
        & Deduplicated & 53 & 94 & 147 \\
        \hline
        
        \multirow{7}{*}{Files} & Total & 8,101,457 & 14,894,317 & 22,995,774  \\
        & Filtered & 7,375,317 & 14,698,938 & 22,074,255\\
        & Deduplicated & 5,101,457 & 10,418,609 & 15,520,066 \\
        & Code & 4,128,813 & 8,380,496 & 12,509,309 \\
        & Test  & 972,644 & 2,038,113 & 3,010,757 \\
        & File pairs & 412,881 & 743,882 & 1,156,763 \\
        & Training & 4,688,576 & 9,674,727 & 14,363,303 \\
        \hline
        \end{tabular}
    \end{center}
}
\end{table}
}

\section{Dataset}

This section describes dataset preparation for both training and evaluating \tool. \Cref{tab:data} provides high-level statistics pertaining to data collection and filtering.

\subsection{Data Collection}
\label{subsec:data}

We use the GitHub API~\cite{github-api} to mine Python and Java repositories that have at least 10 stars and have new commits after January 1st, 2020. 
Following \cite{allamanis2019adverse} and \cite{lopes2017dejavu}, we also remove forks, to prevent data duplication.
This results in a total of 148,605 Python and 49,125 Java repositories with a total of $\sim$23M files (about 280 GB).  We randomly split this into train and test set, ensuring that the test set includes 500 repositories for Python and Java each. 

\subsection{Training Data Preparation}
\label{subsec:preprocessing}

We first remove all non-source code files (e.g., configuration and README files) to ensure that the model is trained on source code only. We then apply a series of filters in accordance with CodeParrot's standards \cite{CodeParrot} to minimize noise from our training signal. This includes removing files that are larger than 1MB, as well as files with any lines longer than 1000 characters; an average line length of $>$100 characters; more than 25\% non-alphanumeric characters, and indicators of being automatically generated. This removes 9\% of both Python and Java files. We deduplicate the files by checking each file's md5 hash against all other files in our corpus. This removes approximately 30\% of both Python and Java files.

We extract code-test file pairs from this data using a combination of exact and fuzzy match heuristics. Given a code file with the name \texttt{<CFN>}, we first search for test files that have the pattern \texttt{test\_<CFN>}, \texttt{<CFN>\_test}, \texttt{<CFN>Test} or \texttt{Test<CFN>}. If no matches are found, we perform a fuzzy string match~\cite{thefuzz} between code and test file names, and group them as a pair if they achieve a similarity score greater than $0.85$. If multiple matches are found, we keep the pair with the highest score.

Following file pair extraction, we prepare our training data by replacing the code and test files with a new file that concatenates the contents of the code file and the test file, separating them with a unique \texttt{<|codetestpair|>} token. This ensures that the model learns the mapping between code and test files from the pretraining signal. Note that we always combine these files starting with the code, so the model (which operates left-to-right) only benefits from this pairing information when generating the test. We additionally include all the other code and test files for which we did not find pairs in our training data, which results in 4.7M Python files and 9.7 Java files. We include these unmatched files to maximize the amount of data the model can learn from. \Cref{fig:code-test-alignment} summarizes the distribution of files in the training data along with sample code snippets for each type of file.

\noindent\textbf{Distribution of files and file pairs: }
Figure~\ref{fig:files} summarizes the distribution of files in projects with respect to their star count. We observe a decreasing trend
in not just the number of code files and test files, but also the file pairs. 
Upon manual inspection of a few randomly selected projects, we find that popular projects with a high star count tend to be better-tested, in line with prior literature~\cite{testingpracicesstudy, souzatestingcontribution}. Note that we normalize
the plot to help illustrate trends by aggregating projects
in buckets based on percentiles, after sorting them based on stars.
The data distribution varies between Python and Java: Python has approximately 3x more projects than Java, but Java has roughly twice as many code-test file pairs.

\begin{figure}
    \centering
    \vspace{-2mm}
    \includegraphics[width=\linewidth,trim={1.5cm 5.5cm 10cm 3.5cm}, clip]{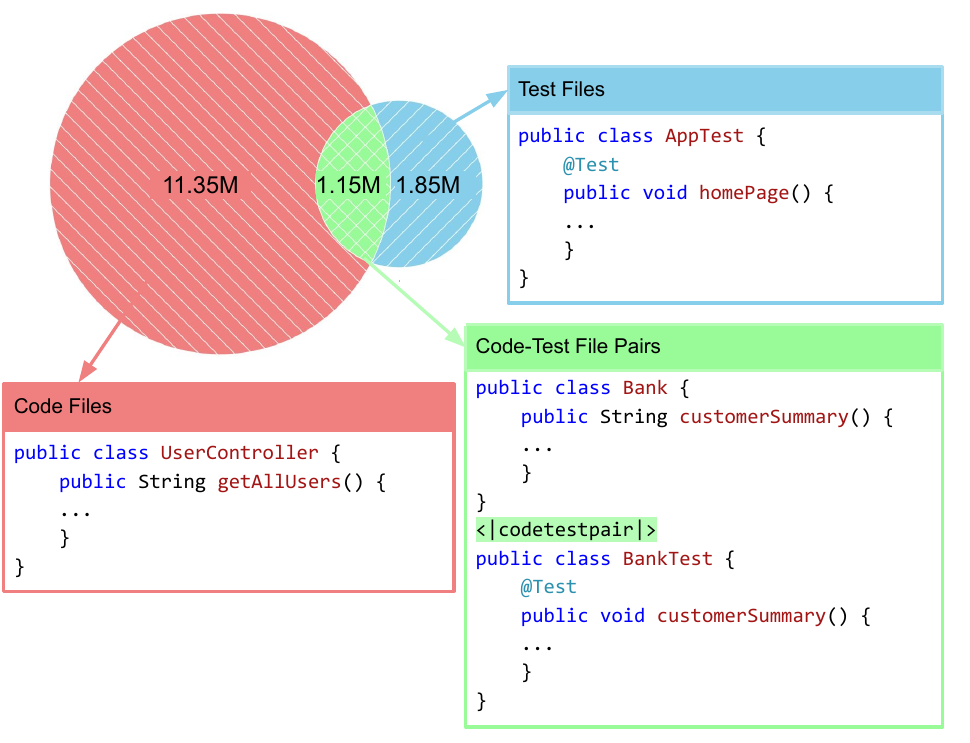}

    \caption{Distribution of files with sample code snippets}
    \label{fig:code-test-alignment}
    \vspace{-2mm}
\end{figure}
  
  \begin{figure}
  \centering
    \includegraphics[width=0.45\textwidth]{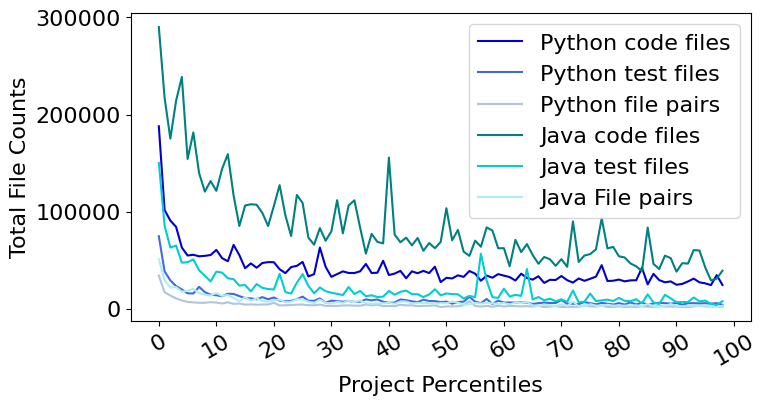}
    \caption{Distribution of files in projects sorted by GitHub stars, normalized by percentiles}
    \label{fig:files}
  \end{figure}
  
\subsection{Test Data Preparation and Execution Setup}
\label{sec:execution-setup}
To prepare our test data, we first excluded all projects without code-test file pairs. This resulted in a total of 97 Java and 152 Python projects. We  then attempted to set up all projects for automated test execution. 

\noindent{\textbf{Execution Setup for Java}: Projects may use different Java versions (which include Java 8, 11, 14, and 17) and build systems (mostly Maven and Gradle). We manually set up Docker images for each combination. We then attempted to execute the build commands for each project in a container from each image. We successfully built 54 out of the 97 Java projects, containing 61 code-test file pairs.

\noindent{\textbf{Execution Setup for Python}}: We manually set up Docker containers for Python 3.8 and 3.10 with the \texttt{pytest} framework and attempted to run the build commands for each project until the build was successful. We successfully built 41 of the 152 Python projects, containing 1080 code-test file-pairs.

We further discarded all \emph{pairs} within these projects with only a single code method or a single test method to ensure that code-test file-pairs in our test set correspond to nontrivial test suites. We additionally require the Java and Python projects to be compatible with the \texttt{Jacoco} and \texttt{coverage} libraries respectively. This leaves a total of 27 code-test file pairs across 26 unique Java projects and 517 code-test file pairs across 26 unique Python projects. In Python, we randomly sampled up to 10 file pairs per project to reduce the bias towards large projects (the top two projects account for 346 tests) leading to a final set of 123 file pairs across 26 unique Python projects. Note that we reuse these Docker containers in our testing framework (See \Cref{sec:testingframework}).

\section{\tool}
\label{sec:training}

This section describes the details for preparing the input, pretraining \tool and generating the outputs.

\subsection{Input Representation for Pretraining \tool}

We use the corpus of 14M Java and Python files that we prepared for the pretraining of our model (see \Cref{subsec:data}). We first train a subword tokenizer \cite{kudo2018subword} using the SentencePiece~\cite{spm} toolkit with a vocabulary size of 64K tokens. The tokenizer is trained over 3 GB of data using ten random lines sampled from each file. We then tokenize our input files into a binary format used to efficiently stream data during training.

\noindent\textbf{Analysing the distribution of tokens:}
Language models are typically constrained in the amount of text they fit in their context window. Most current code generation models use a context window of up to 2,048 tokens \cite{CodeGen, XuPolyCoder}.\footnote{The average length of a token depends on the vocabulary and dataset, but can typically be assumed to be around 3 characters.} Our analysis on the distribution of tokens, visualized in \Cref{fig:sum_tokens}, showed that this only covers 35\% of the total number of file pairs. As such, while it may be appropriate for a (slight) majority of individual files, it would not allow our model to leverage the code file's context while predicting text in the test file. This is a significant limitation since we want to train the model to use the context from the code file when generating tests. 

Further analysis showed that approximately 82\% of all file pairs for Java and Python have fewer than 8,192 tokens. Since the cost of the attention operation increases quadratically with the context length, we choose this cutoff to balance training cost and benefit. Therefore, we chose to train a model with a longer context window of 8192 tokens to accommodate an additional $\sim$550K file pairs. Note that this does not lead to any samples being discarded; pairs with more tokens will simply be (randomly) chunked by the training toolkit.

    \begin{figure}
     \centering
     \includegraphics[width=0.45\textwidth]{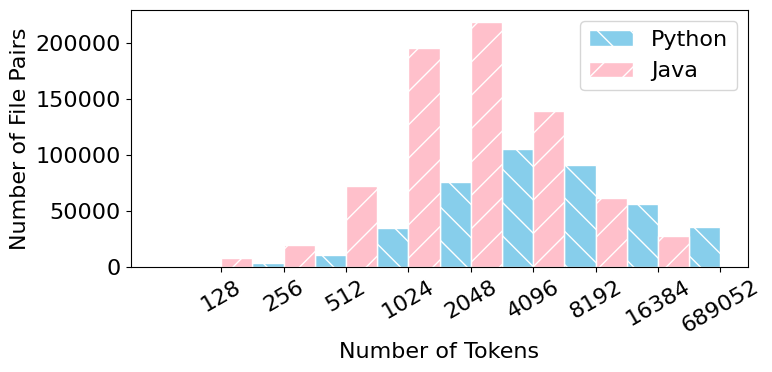}
    \caption{Distribution of file pair tokens}
    \label{fig:sum_tokens}
  \end{figure}

\subsection{Model and Training Details}
We determined the model size based on our cloud compute budget of \$20,000 and the amount of available training data, based on the Chinchilla scaling laws~\cite{chinchilla}, which suggest that the training loss for a fixed compute budget can be minimized (lower is better) by training a model with ca. (and no fewer than) 20 times as many tokens as it has parameters. Based on preliminary runs, we determined the appropriate model size to be 2.7 (non-embedding) parameters, a common size for medium to large language models \cite{XuPolyCoder, CodeGen}, which we therefore aimed to train with at least 54B tokens. This model architecture consists of a 2,560-dimensional, 32 layer Transformer model with a context window of 8,192 tokens. We trained the model with a batch size of 256 sequences, which corresponds to $\sim$2M tokens. We use the GPT-NeoX toolkit~\cite{neox} to train the model efficiently with 8 Nvidia A100 80GB GPUs on a single machine on the Google Cloud Platform. We trained the model for 28.5K steps, for a total of nearly 60B tokens, across 18 days, thus averaging roughly ~1,583 steps per day\footnote{\change{We further trained the model to 35.3K steps, thanks to an additional grant received for \$5000, after the paper was published. This latest checkpoint is now available on HuggingFace. Please see \url{https://github.com/RaoNikitha/CAT-LM} for more details on usage. Note that the numbers reported in this paper make use of the older checkpoint (28.5K steps), and may not match the numbers from the newer public checkpoint (35.3K steps).}} %While the training and validation loss is down to 0.4023 and 0.4368 respectively at the final step, the validation loss still decreases, which means that the model can benefit from even further training, which we did not have the budget for.
We note that this training duration is much shorter than many popular models \cite{CodeGen, touvron2023llama};\footnote{The ``Chinchilla" optimum does not focus on maximizing the performance for a given model size, only for a total compute budget.} the model could thus be improved substantially with further training. The final model is named \tool as it is trained on aligned \textbf{C}ode \textbf{A}nd \textbf{T}ests.

\subsection{Prompting \tool to generate outputs:} 
Since \tool has been trained using a left-to-right autogressive pretraining signal, it can be prompted to generate some code based on the preceding context. In our case, we task it to either generate an entire test method 
given the preceding test (and usually, code) file context, or generating a line to complete the test method (given the same). We prompt \tool with the inputs for each task, both with and without code context, and sample 10 outputs from \tool with a ``temperature" of 0.2, which encourages generating different, but highly plausible (to the model) outputs. Sampling multiple outputs is relatively inexpensive given the size of a method compared to the context size, and allows the model to efficiently generate multiple methods from an encoded context. We can then filter out tests that do not compile, lack asserts, or fail (since we are generating behavioral tests), by executing them in the test framework. We prepare the outputs for execution by adding the generated test method to its respective position in the baseline test files, without making any changes to the other tests in the file.
\section{Experimental Setup}
\label{sec:evaluation}

We describe the setup for evaluating \tool across both tasks outlined in \Cref{sec:tasks}, namely test method generation, and test completion. 

\subsection{Test Method Generation} 

The test method generation task involves three different cases: generating the first test, the final test, and an extra test in a test suite (see \Cref{sec:tasks}). We evaluate \tool on test method generation both with code context and, as an ablation, without code context.

\subsubsection{Baseline Models} CodeGen is a family of Transformer-based LLMs trained auto-regressively (left-to-right) \cite{CodeGen}. Pretrained CodeGen models are available in a wide range of sizes, including 350M, 2.7B, 6.1B and 16.1B parameters. These models were trained on three different datasets, starting with a large, predominantly English corpus, followed by a multi-lingual programming language corpus (incl. Java and Python), and concluding with fine-tuning on Python data only. The largest model trained this way is competitive with Codex~\cite{Chen2021Codex} on a Python benchmark~\cite{CodeGen}.

For our evaluation, we compare with CodeGen-2.7B-multi, which is comparable in size to our model and trained on multiple programming languages, like our own. We also consider CodeGen-16B-multi (with 16B parameters, ca. 6 times larger than \tool) which is the largest available model trained on multiple programming languages. For all Python tasks, we also compare against CodeGen-2.7B-mono and CodeGen-16B-mono, variants of the aforementioned models fine-tuned on only Python code for an additional 150k training steps.

We also compare the performance of \tool with StarCoder~\cite{starcoder}, which is a 15.5B parameter model trained on over 80 programming languages, including Java and Python, from The Stack (v1.2). StarCoder has a context window of $8,192$ tokens. It was trained using the Fill-in-the-Middle objective~\cite{FiMOpenAI} on 1 trillion tokens of code, using the sample approach of randomizing the document order as CodeGen.

\subsubsection{Lexical Metrics}
\label{sec:lexmetrics}

Although our goal is not to exactly replicate the human-written tests, we provide measures of the \emph{lexical} similarity between the generated tests and their real-world counterparts as indicators of their realism. Generated tests that frequently overlap in their phrasing with ground-truth tests are likely to be similar in structure and thus relatively easy to read for developers.
Specifically, we report both the rate of exact matches and several measures of approximate similarity, including  ROUGE~\cite{lin-2004-rouge} (longest overlapping subsequence of tokens) and CodeBLEU~\cite{RenCodeBleu20} score ($n$-gram overlap that takes into account code AST and dataflow graph). We only report lexical metrics for our first test and last test settings, as there is no ground truth to compare against in our extra test setting. These metrics have been used extensively in prior work on code generation and test completion~\cite{NieETAL22Teco, DeepCommentGenESE, LeClair19NeuralSubroutines, WangCodeT5}.

\subsubsection{Runtime Metrics} 
\label{sec:runtimemetrics}

We also report runtime metrics that better gauge test utility than the lexical metrics. This includes the number of generated tests that compile, and generated tests that pass the test suite. We also measure coverage of the generated tests. For first and last tests, we compare this with the coverage realized by the corresponding human-written tests. We hope that this work will encourage more widespread adoption of runtime metrics (which are an important part of test utility), as prior work primarily focuses on lexical similarity \cite{DinellaTOGA, WatsonATLAS, NieETAL22Teco}. See Section 2.2 in supplementary material for additional detailed descriptions of all lexical and run-time metrics.

\subsubsection{Preparing Input Context and Baseline Test Files} We use an AST parser on the ground-truth test files to prepare partial tests with which to prompt \tool. For first test generation, we remove all test cases (but not the imports, nor any other setup code that precedes the first test); for last test generation, we leave all but the final test method, and for final test generation we only remove code after the last test. We then concatenate the code context to the test context using our delimiter token for the `with code context' condition.

We additionally obtain coverage with the original, human-written test files under the same conditions, keeping only the first or all tests as baselines for first and last test prediction respectively. Note that there is no baseline for the extra test generation task. See Section 1 for in supplementary material for coverage distribution of human-written tests.

\subsubsection{Testing Framework}
\label{sec:testingframework} 

We evaluate the quality of the generated tests using the containers that we setup to execute projects in \Cref{sec:execution-setup}. We insert the generated test into the original test file, execute the respective project's setup commands and check for errors, recording the number of generated tests that compile and pass the test suite (see \Cref{sec:runtimemetrics}). If the generated test compiles successfully (or, for Python, is free of import or syntax errors), we run the test suite and record whether the generated test passed or failed. We compute code coverage for all passing tests, contrasting this with the coverage achieved by the human-written test cases (when available) as baselines.

\begin{table}[t]
{\small
    \begin{center}
    \centering
    \caption{ Baseline coverage for human written tests over the given number of file pairs.}
    \label{tab:baselinecov}
    \begin{tabular}{l|l|r|r}
    \toprule
    \textbf{PL} & \textbf{Case} & \textbf{Cov Imp \%} & \textbf{\# File Pairs} \\
    \midrule
    \multirow{3}{*}{Python} & First test & 59.3\% & 112 \\
    & Last test & 5.0\% & 93 \\ 
    & Extra test & 0.0\% & 123 \\ \midrule
    \multirow{3}{*}{Java} & First test & 50.5\% & 27 \\ 
    & Last test & 5.3\% & 18 \\
    & Extra test & 0.0\% & 27 \\
    \bottomrule
    \end{tabular}
    \end{center}
}
\end{table}

\subsection{Test Completion}

Recall the test completion task involves generating a single line in a given test method, given the test's previous lines. We perform our evaluation for test completion under two conditions, with code context and without code context. 

\subsubsection{Baseline Model} We compare against TeCo~\cite{NieETAL22Teco}, a state of the art baseline on test statement completion that has outperformed many existing models, including CodeT5~\cite{WangCodeT5}, CodeGPT~\cite{CodeGPT} and TOGA~\cite{DinellaTOGA}. TeCo~\cite{NieETAL22Teco} is a encoder-decoder transformer model based on the CodeT5 architecture~\cite{WangCodeT5}. TeCo takes the test method signature, prior statements in the test, the method under test, the variable types, absent types and method setup and teardown as input.

Initially, we intended to compare \tool against TeCo on our test set. However, TeCo performs extensive filtering including requiring JUnit, Maven, well-named tests, a one-to-one mapping between test and method under test, and no if statements or non-sequential control flow in the test method. We thus compared \tool against TeCo for 1000 randomly sampled statements from their test set. 

\subsubsection{Metrics} We compare \tool against TeCo across all lexical metrics (outlined in \Cref{sec:lexmetrics}).

\begin{figure}[t]
  \centering
  \vspace{-3mm}
  \begin{subfigure}{0.45\textwidth}
    \includegraphics[width=\textwidth]{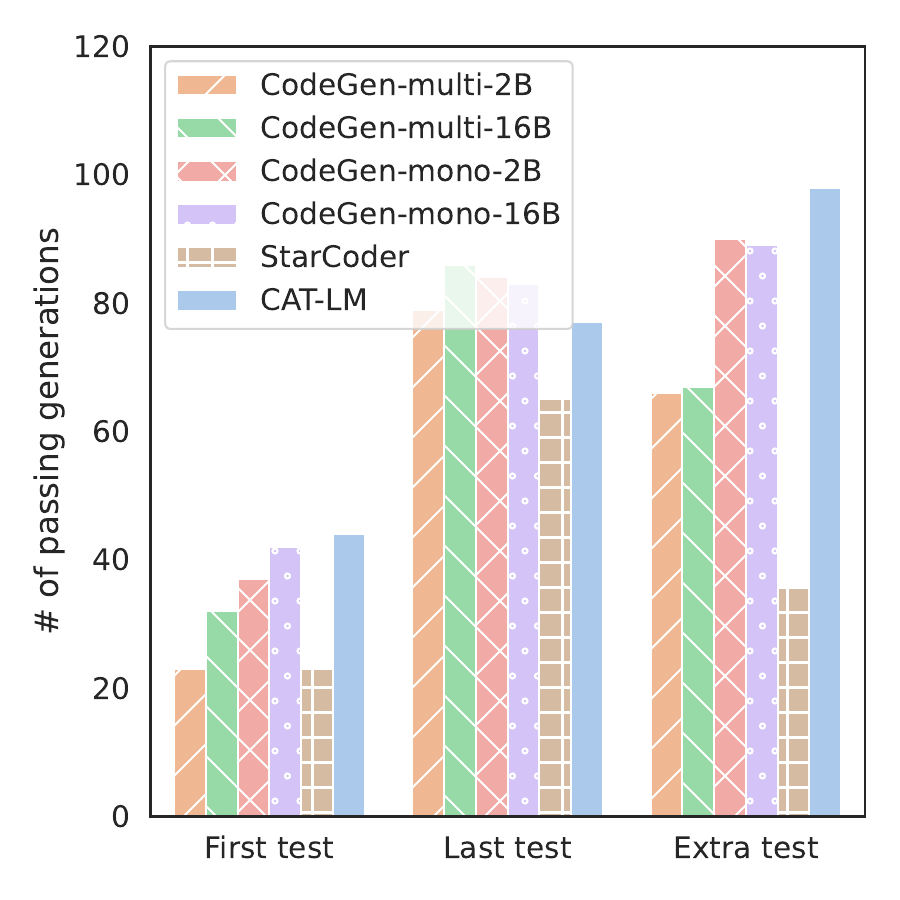}
    \caption{Python.}
    \label{fig:pythonpass}
  \end{subfigure}
  \hfill
  \begin{subfigure}{0.45\textwidth}
    \includegraphics[width=\textwidth]{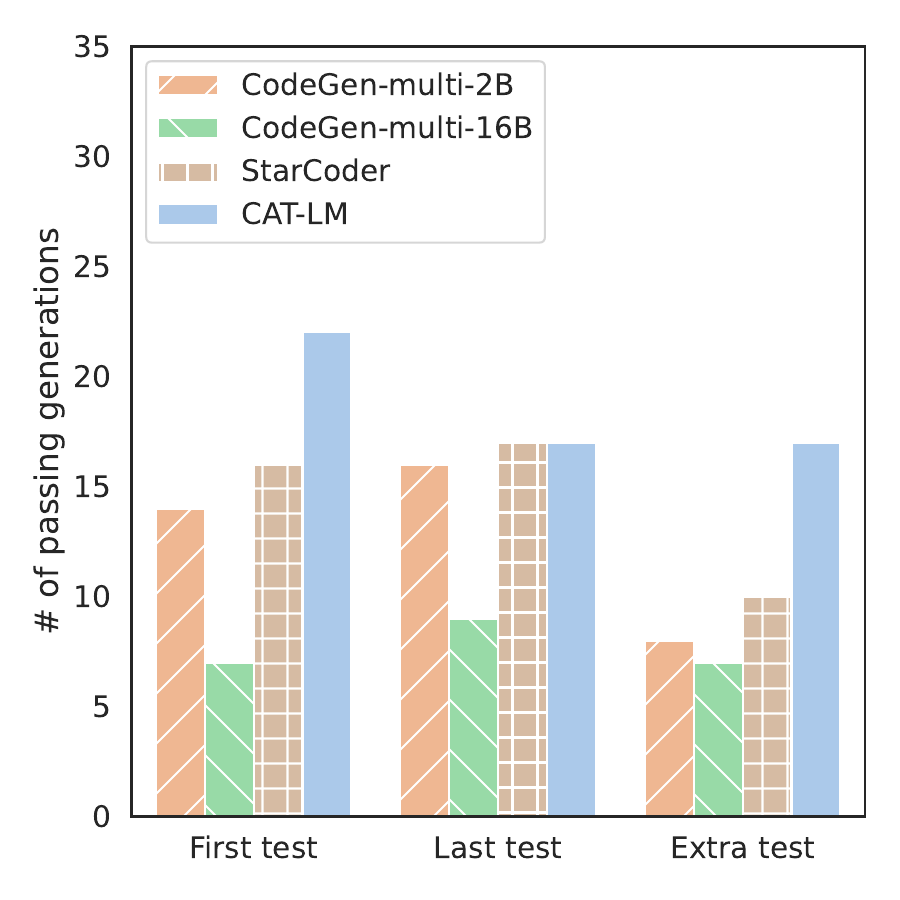}
    \caption{Java.}
    \label{fig:javapass}
  \end{subfigure}
  \caption{Passing tests by model for Python and Java.}
  \label{fig:passrates}
\end{figure}

\begin{figure}[t]
  \centering
  \vspace{-9mm} 
  \begin{subfigure}{0.45\textwidth}
    \includegraphics[width=\textwidth]{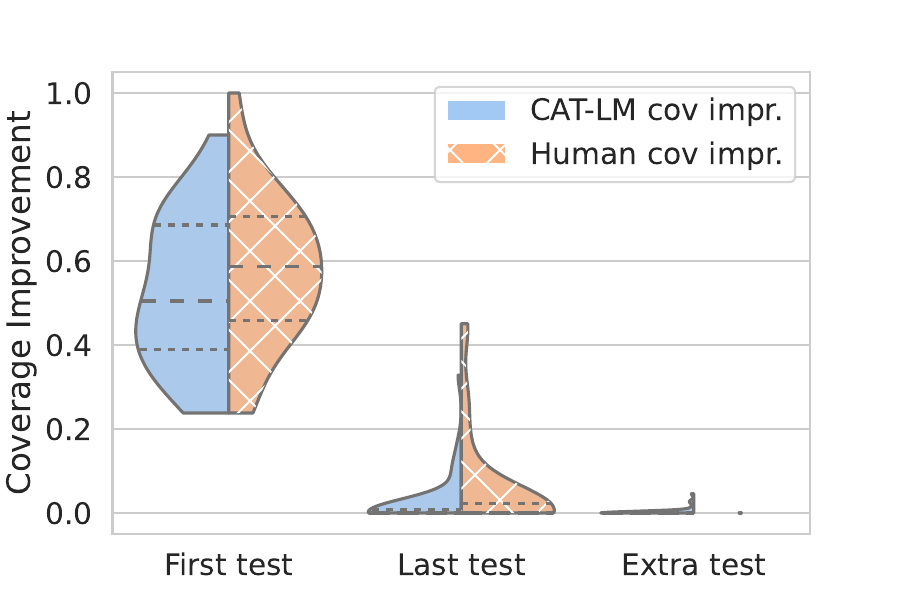}
    \caption{Coverage improvement of our model vs humans for Python.}
    \label{fig:pythonpass}
  \end{subfigure}
  \hfill
  \begin{subfigure}{0.45\textwidth}
    \includegraphics[width=\textwidth]{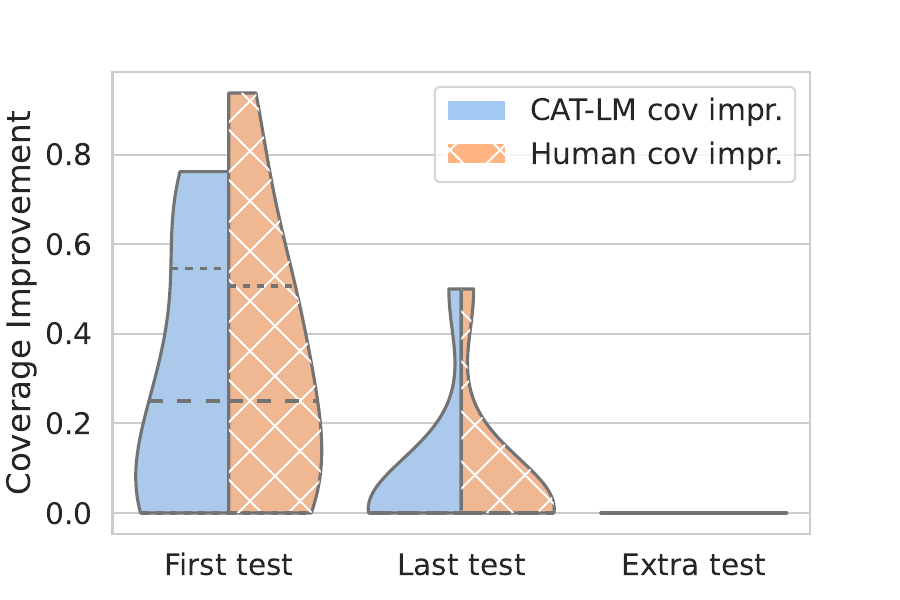}
    \caption{Coverage improvement of our model vs humans for Java.}
    \label{fig:javapass}
  \end{subfigure}
  \caption{ Coverage improvement of our model vs humans for different languages.}
  \label{fig:covdist}
\end{figure}

\section{Evaluation}

We evaluate \tool's ability to generate valid tests that achieve coverage, comparing against state of the art baselines for both code generation and test completion. Additional results can be found in the supplementary material.

\subsection{Test Method Generation}
\subsubsection{Pass Rate}

\Cref{fig:passrates} shows the number of passing tests generated by each model for Python and Java. Note that these are absolute numbers, out of a different total for each setting.\footnote{The denominator for each group is the number of file pairs shown in \Cref{tab:baselinecov} multiplied by 10, the number of samples per context.} 

\tool outperforms StarCoder and all CodeGen models, including ones that are much larger and language-specific in most settings. 
For Python, all models perform worst in the first test setting, where they have the least context to build on. Nonetheless, equipped with the context of the corresponding code file, our model generates substantially more passing tests than StarCoder (with 15.5B parameters) and the multilingual CodeGen baselines (trained with far more tokens) in both first and extra test setting. Only in the last-test settings do some of the models compete with ours, though we note that their performance may be inflated as the models may have seen the files in our test set during training (the test set explicitly omits files seen by \tool during training).
For Java, we find that \tool generates more passing tests than StarCoder and the two multilingual CodeGen models (no Java-only model exists). The difference is most pronounced in the extra test setting, where \tool generates nearly twice as many passing tests compared to StarCoder and the CodeGen baseline models.
Overall, despite being undertrained, \tool generates more number of passing tests on average across all settings. Both StarCoder and the CodeGen models don't show significant gains with more parameters or longer contexts (StarCoder can use $8,192$ tokens), highlighting that training with code context is important.

\begin{table*}
{\small
\begin{center}
\centering
\caption{Lexical and runtime metrics performance comparison of the models on the held-out test set for Java and Python. CodeGen refers to CodeGen-multi for Java and CodeGen-mono for Python results. We only report lexical metrics for our first test and last test settings, as there is no gold test to compare against in our extra test setting.}
\label{tab:test-gen-results}
\begin{tabular}{@{}l|rrr|rr|rrr|rr@{}}
\toprule
 & \multicolumn{5}{c|}{\textbf{Java}} & \multicolumn{5}{c}{\textbf{Python }} \\\toprule
 
 & \multicolumn{3}{c|}{\textbf{Lexical Metrics}} & \multicolumn{2}{c|}{\textbf{Runtime Metrics}} & \multicolumn{3}{c|}{\textbf{Lexical Metrics}} & \multicolumn{2}{c}{\textbf{Runtime Metrics}} \\\toprule

\textbf{Model} & \textbf{CodeBLEU} & \textbf{XMatch} & \textbf{Rouge} &  \textbf{Compile} & \textbf{Pass} & \textbf{CodeBLEU} & \textbf{XMatch} & \textbf{Rouge} & \textbf{Compile} & \textbf{Pass} \\ \midrule

\multicolumn{11}{c}{\textbf{First Test (Total: Java = 270, Python = 1120)}} \\\midrule
\tool w Context &  41.4\% & \textbf{15.4\%} & 60.9\% & \textbf{50} & \textbf{22} & 21.0\% & 0.3\% & \textbf{39.4\%} & \textbf{384} & \textbf{44} \\
\tool w/o Context & 37.5\% & \textbf{15.4\%} & 56.5\% & 9 & 9 & 17.7\% & 0.4\% & 30.2\% & 236 & 31 \\
Codegen-2B & 35.5\% & 7.7\% & 56.8\% & 24 & 14 & 18.2\% & 0.0\% & 30.9\% & 259 & 37 \\
Codegen-16B & 42.2\% & 7.7\% & 61.8\% & 25 & 7 & 20.8\% & 0.3\% & 35.1\% & 361 & 42 \\
StarCoder & \textbf{44.6\%} & 10.9\% & \textbf{62.2\%} & 28 & 16 & \textbf{24.0\%} & \textbf{1.8\%} & 38.8\% & 269 & 23 \\
\midrule\multicolumn{11}{c}{\textbf{Last Test (Total: Java = 180, Python = 930)}} \\\midrule
\tool w Context & 55.4\% & 20.8\% & 70.8\% & \textbf{54} & \textbf{17} & \textbf{38.3\%} & \textbf{4.8\%} & \textbf{54.9\%} & 335 & 77 \\
\tool w/o Context & 53.6\% & 20.8\% & 68.9\% & 33 & 14 & 33.2\% & 1.4\% & 51.9\% & \textbf{350} & 79 \\
Codegen-2B & 51.7\% & 13.0\% & 69.2\% & 43 & 16 & 36.3\% & 2.2\% & 53.2\% & 326 & \textbf{84} \\
Codegen-16B & 56.5\% & 14.3\% & \textbf{70.9\%} & 24 & 9 & 37.9\% & 3.4\% & 54.0\% & 349 & 83 \\
StarCoder & \textbf{56.9}\% & \textbf{21.0\%} & 69.9\% & 34 & \textbf{17} & 37.6\% & 4.2\% & 54.5\% & 227 & 65 \\
\midrule\multicolumn{11}{c}{\textbf{Extra Test (Total: Java = 270, Python = 1230)}} \\\midrule
\tool w Context & -- & -- & -- & \textbf{41} & 17 & -- & -- & -- & 380 & 98 \\
\tool w/o Context & -- & -- & -- & 29 & \textbf{20} & -- & -- & -- & \textbf{425} & \textbf{104} \\
Codegen-2B & -- & -- & -- & 17 & 8 & -- & -- & -- & 376 & 90 \\
Codegen-16B & -- & -- & -- & 15 & 7 & -- & -- & -- & 384 & 89 \\
StarCoder & -- & -- & -- & 17 & 10 & -- & -- & -- & 269 & 36 \\
\bottomrule
\end{tabular}
\end{center}}
\vspace{-4mm}
\end{table*}

\subsubsection{Coverage}
\Cref{fig:covdist} shows the coverage distribution of \tool, contrasted with that of the human-written tests. For both the first test and last test settings, our model performs mostly comparably to humans, with both distributions having approximately the same median and quartile ranges. The extra test task is clearly especially hard: while our model was able to generate many tests in this setting (\Cref{fig:passrates}), these rarely translate into \emph{additional} coverage, beyond what is provided by the rest of the test suite, in part because most of the developer-written test suites in our dataset already have high code coverage (average coverage of 78.6\% for Java and 81.6\% for Python), and may have no need for additional tests. \Cref{tab:baselinecov} shows the average human coverage improvement for the first and last test added to a test suite. Note that the average is significantly lower for last test, as baseline coverage is already high for this mode (74.7\% for Java and 76.1\% for Python).

We note that we could not compute coverage for all the file pairs in each setting. We excluded file pairs with only one test from our last test setting to differentiate it from our first test setting. For the first test setting, some baseline files were missing helper methods between the first test and last test in the file, preventing us from computing coverage.

\subsubsection{Lexical Similarity}
\Cref{tab:test-gen-results} shows the lexical similarity metrics results relative to the human-written tests for \tool, both with and without context, along with StarCoder and CodeGen baselines. \tool reports high lexical similarity scores when leveraging code context, typically at or above the level of the other best model, StarCoder (with 15B parameters). This effect is consistent across first and last test generation.

\subsubsection{Impact of Code Context}

As is expected, \tool heavily benefits from the presence of code context. When it is queried without this context, its performance on lexical metrics tends to drop to below the level of CodeGen-2B, which matches it in size but was trained with more tokens. The differences in lexical metric performance are sometimes quite pronounced, with up to a 9.2\% increase in Rouge score and up to a 5.1\% increase in CodeBLEU score.

In terms of runtime metrics, code context mainly helps on the first and last test prediction task, with especially large gains on the former. Context does not seem to help generate more passing tests in the extra test setting. This may be in part because the test suite is already comprehensive, so the model can infer most of the information it needs about the code under test from the tests. It may also be due to the test suites often being (nearly) complete in this setting, so that generating additional tests that pass (but yield no meaningful coverage) is relatively straightforward (e.g., by copying an existing test \Cref{sec:casestudy}).
Overall, these results support our core hypothesis that models of code should consider the relationship between code and test files to generate meaningful tests.

\subsubsection{Other Runtime Metrics} 
\Cref{tab:test-gen-results} also shows a comparison between \tool and StarCoder and CodeGen baselines for all runtime metrics. 
\tool outperforms both StarCoder and the CodeGen baselines in both Python in Java across compiling and passing generations, with \tool typically generating the most samples that compile and pass.  
The one setting where the CodeGen baselines perform slightly better is in generating more last tests that pass for Python. However, the compile rate of these CodeGen generated tests is significantly lower than those generated by \tool. 
We note that CodeGen’s performance may be inflated in the last test setting, as it may have seen the files from the test set during training.

\begin{highlight}
 \textbf{\tool outperforms StarCoder and CodeGen} for both Python and Java, \textbf{generating more passing tests} on average across all settings. We find that \textbf{code context improves performance} across most settings in terms of both lexical and runtime metrics.
\end{highlight}

\begin{table}[ht]
{\small
\begin{center}
\centering
\caption{Comparison of \tool and TeCo on 1000 randomly sampled statements in their test set.}
\label{tab:lexicalteco}
\begin{tabular}{l|r|r|r}
\toprule
\textbf{Model} & \textbf{CodeBLEU} & \textbf{XMatch} & \textbf{Rouge} \\
\midrule
\tool w/ Context & \textbf{67.1\%} & \textbf{50.4\%} & \textbf{82.8\%} \\
\tool w/o Context & 65.9\% & 48.9\% & 82.2\% \\
TeCo & 26.7\% & 13.8\% & 60.2\% \\
\bottomrule
\end{tabular}
\end{center}
\vspace{-4mm}
}
\end{table}
\subsection{Test Completion}

For test completion (see \Cref{sec:testcompletion} for task definition), we compare \tool against TeCo \cite{NieETAL22Teco} on the lexical metrics outlined in \Cref{sec:lexmetrics}.
Specifically, we sample 1000 statements at random from across the test set released by the authors of TeCo, on which we obtain similar performance with TeCo to those reported in the original paper. 
\Cref{tab:lexicalteco} shows the results. 
\tool outperforms TeCo across all lexical metrics, with a 36.6\% increase in exact match, 22.6\% increase in ROUGE and 40.4\% increase in CodeBLEU score. Even prompting \tool with just the test context (i.e., without the code context) yields substantially better results than TeCo. This underscores that providing the entire test file prior to the statement being completed as context, rather than just the setup methods, is helpful for models to reason about what is being tested.

In contrast to the test generation task, code context only slightly helps \tool in this setting, with an increase in CodeBLEU score of 1.2\% and increase in exact match accuracy of 1.5\%. Apparently, many individual statements in test cases can be completed relatively easily based on patterns found in the test file, without considering the code under tests. This suggests that statement completion is significantly less context-intensive than whole-test case generation. We therefore argue that entire test generation is a more appropriate task for assessing models trained for test generation.

\begin{highlight}
\textbf{\tool outperforms TeCo across all lexical metrics}, with a 40.4\% improvement in CodeBLEU score and 36.6\% improvement in exact match accuracy. We find that \textbf{context only slightly helps} with test statement prediction, indicating that test completion can largely be done without the code under test, in contrast to entire test generation.
\end{highlight}

See Section 3 and 4 in the supplementary material for additional results on all tasks.

\begin{figure}[!th]
\centering
\begin{minipage}{0.45\textwidth}
\begin{lstlisting}[caption=CAT-LM first test generation, label=lst:cat-lmfirst]
@Test
public void testCustomerSummary() {
    Bank bank = new Bank();
    assertEquals("Customer Summary", "Customer Summary", bank.customerSummary());
}
\end{lstlisting}
\vspace{-3mm}

\begin{lstlisting}[caption=CodeGen first test generation, label=lst:codegenfirst]
@Test
public void testBank() {
    Bank bank = new Bank();
    assertEquals(0, bank.getBalance(), DOUBLE_DELTA);
    bank.deposit(100);
    assertEquals(100, bank.getBalance(), DOUBLE_DELTA);
    bank.withdraw(50);
    assertEquals(50, bank.getBalance(), DOUBLE_DELTA);
}
\end{lstlisting}
\vspace{-3mm}

\begin{lstlisting}[caption=GPT-4 first test generation, label=lst:gpt4first]
@Before
public void setUp() {
    bank = new Bank();
}

@Test
public void testAddCustomerAndCustomerSummary() {
    Customer customer1 = new Customer("John Doe");
    bank.addCustomer(customer1);

    String expectedSummary = "Customer Summary\n - John Doe (0 accounts)";
    String actualSummary = bank.customerSummary();

    assertEquals(expectedSummary, actualSummary);
}
\end{lstlisting}
\vspace{-3mm}

\begin{lstlisting}[caption=EvoSuite first test generation, label=lst:evosuitefirst]
@Test(timeout = 4000)
public void test0() throws Throwable  {
    Bank bank0 = new Bank();
    Customer customer0 = new Customer("v\"PD");
    bank0.addCustomer(customer0);
    Account account0 = new Account(0);
    account0.deposit(148.3628547);
    customer0.openAccount(account0);
    double double0 = bank0.totalInterestPaid();
    assertEquals(0.14836285470000002, double0, 0.01);
}
\end{lstlisting}
\vspace{-3mm}
\end{minipage}
\caption{ Example first tests generated by \tool, CodeGen, GPT-4,  and EvoSuite. \tool and GPT-4 both generate realistic and readable tests; EvoSuite struggles with poor naming conventions and unrealistic tests. CodeGen generates readable test cases, but hallucinates methods in the code under test. See Section 5 in supplementary material for additional examples.}
\label{fig:toolcompletions}
\end{figure}

\subsection{Qualitative Comparisons}
\label{sec:casestudy}
Finally, we conduct a small-scale qualitative case-study of tests generated by \tool, CodeGen-2B-multi \cite{CodeGen}, GPT-4~\cite{openai2023gpt4} and EvoSuite~\cite{FraserEvoSuite}. GPT-4 is a vastly larger language model than ours, trained with an undisclosed budget by OpenAI. EvoSuite is a popular test generation tool for Java based on evolutionary algorithms.

We analyze a randomly sampled passing generation from \tool in contrast to the tests generated by the other tools in the same context across each our three settings (first test, last test and extra test). The tests here are generated for a \texttt{Bank} class, which includes methods to add a customer, open an account and print a summary of all accounts and customers. Our goal is to better understand the benefits and drawbacks of each tool's generated tests. Specifically, we look for characteristics of high quality tests, such as meaningful method and variable names, proper invocation of the method under test and high quality assertions. We mainly discuss the first generated test here; for our full set of examples, see Supplemental Materials.

\noindent\textbf{\tool:} Listing \ref{lst:cat-lmfirst} shows the first test generation by \tool. The name of the test is informative, along with its variables. It also follows unit testing conventions of testing one specific method in the \texttt{Bank} class. This is consistent across the examples for last test and extra test. However, for our extra test example, \tool copied the previous test and changed the name of the test method, not testing new functionality.

\noindent\textbf{CodeGen:} in Listing \ref{lst:codegenfirst}, the test generated by CodeGen is quite readable, semantically correct, and natural looking. However, it uses multiple non-existent methods from the code under test---a phenomenon popularly dubbed ``hallucinating"---since it lacks awareness of \texttt{Bank}'s implementation. StarCoder performs similarly, generating tests that are readable, semantically correct, and natural looking but suffer from hallucinations.

\noindent\textbf{GPT-4:} GPT-4 consistently performs the best of all three tools, generating tests that either are identical to the ground truth or test new functionality that none of the existing tests do. Listing \ref{lst:gpt4first} shows GPT-4's generation for the first test case. Similar to \tool, the GPT-4 generated test has meaningful identifier names and assertions. GPT-4 had similarly good tests for our last test and extra test settings.
However, these results come with several caveats. First, GPT-4 was trained on a very large volume of data, including public code, so it is quite likely that it was trained on our test data and has thus  seen the original tests.\footnote{In fact, a similar caveat applies to CodeGen, which we do outperform.} Second, GPT-4 is a much larger, 
%closed-source 
model, with a training budget orders of magnitude higher than ours. Given our strong performance compared to the (already much more expensive) CodeGen models, we expect that modestly scaling up our training approach could well yield similar or better results.

\noindent\textbf{EvoSuite:} EvoSuite performs the worst in all three settings. Listing \ref{lst:evosuitefirst} shows the EvoSuite completion for the bank class. The generated test uses very poor naming conventions, such as naming the method \texttt{test0}, and each of the variables \texttt{bank0}, \texttt{customer0}, and \texttt{account0}. The deposit amounts do not make logical sense, as they are not rounded to the nearest cent. There is also a timeout of 4000 milliseconds. Such timeouts are highly likely to lead to flaky tests, where this test might pass in one environment and timeout in a different environment. The other generations by EvoSuite, suffer similar problems, including lacking asserts and using spurious exception handling. Due to this lack of proper naming conventions and the use of trivial asserts, it is very difficult to understand what is being tested in EvoSuite's generation.

\begin{highlight}
Both GPT-4 and \tool generate \textbf{high quality tests}, checking for realistic situations with readable asserts. However \tool struggles to generate meaningfully distinct tests in the extra test setting. CodeGen and StarCoder produces highly readable, but incorrect tests. EvoSuite struggles to generate meaningful tests; it uses \textbf{poor naming conventions} and \textbf{spurious exception handling}.
\end{highlight}

\section{Related Work}
\noindent\underline{\textbf{Classical Test Generation:}} Classical test generation techniques employ both black-box and white-box techniques to generate test inputs and test code. Random/fuzzing techniques such as Randoop~\cite{PachecoRandoop}, aflplusplus~\cite{FioraldiAflplusplus} and honggfuzz use coverage to guide generation of test prefixes. Property testing tools such as Korat~\cite{BoyapatiKorat}, QuickCheck~\cite{koenquickcheck} and Hypothesis~\cite{MacIver2019Hypothesis} allow a developer to specify a set of properties and subsequently generates a suite of tests that test the specified properties. PeX~\cite{TillmannPex} and Eclipser~\cite{ChoiEclipser} use dynamic symbolic execution to reason about multiple program paths and generate interesting inputs. The core issue with fuzzing and classical test generation techniques is their reliance on program crashing or exceptional behavior in driving test generation~\cite{DinellaTOGA}, which limits the level of testing they provide. EvoSuite~\cite{FraserEvoSuite} addresses these challenges by using mutation testing to make the generated test suite compact, without losing coverage. However, EvoSuite generates tests that look ``unnatural'', and significantly different from human tests, suffering from both stylistic and readability problems~\cite{BrandtEvoSuiteStudy, DakaEvoSuiteUserStudy, RobertsonEvoSuite}.

\noindent\underline{\textbf{Neural Test Generation:}} More recently, neural test generation methods have been developed to generate more natural and human understandable tests. ConTest\cite{VillmowContest} makes use of a generic transformer model, using the tree representation of code to generate assert statements. ATLAS~\cite{WatsonATLAS}, ReAssert~\cite{White2020ReAssertDL}, AthenaTest~\cite{TufanoAthenaTest} and TOGA~\cite{DinellaTOGA} extend this work by leveraging the transformer architecture for this task. They show that their generated asserts are more natural and preferred by developers when comparing against existing tools such as EvoSuite. TeCo~\cite{NieETAL22Teco} expands the scope of test completion by completing statements in a test, one statement at a time. They leverage execution context and execution information to inform their prediction of the next statement, outperforming TOGA and ATLAS on a range of lexical metrics. While these neural approaches solve many of the readability issues of classical test generation approaches, they focus on generating individual statements in a test, which offers significantly less time saving benefits than generating entire tests.

\noindent\underline{\textbf{Large Language Models of Code:}} Large language models (LLMs) can perform well across many tasks when prompted with instructions and examples~\cite{BrownGPT3, touvron2023llama}. Codex~\cite{Chen2021Codex} is an autoregressive (left to right generation) LLM with 12B parameters, fine-tuned from GPT-3 on 54 million GitHub Python repositories. CodeGen-16B, with which we compare, outperforms this model \cite{CodeGen}. Later, unpublished, iterations of Codex have also been applied to commercial settings, powering GitHub's Copilot~\cite{copilot}. TestPilot~\cite{schafer2023adaptive} uses Codex to generate unit tests. However, it requires significant volumes of documentation as input, which is often not available for open-source projects.
While all of these models perform well at generating code, they are relatively poor (for their size) at generating \emph{tests} for the code. These models are typically trained on a randomly shuffled corpus of entire files, and thus do not learn the alignment of tests to the code under test. We pretrained a comparatively small language model on a much more modest budget that explicitly learns to align code and the corresponding test files, which yields substantially better performance than modestly larger classically trained models.

\section{Conclusion}
We develop \tool, a GPT-style language model with 2.7 Billion parameters that was pretrained using a novel signal that explicitly considers the mapping between code and test files when available. We elect to use a larger context window of 8,192 tokens, 4x more than typical code generation models, to ensure that code context is available when generating tests. We evaluate \tool on both test method generation and test completion, with \tool outperforming CodeGen, StarCoder, and TeCo state-of-the-art baselines, even with CodeGen and StarCoder baselines significantly larger training budgets and model sizes. We show that adding the additional context helps \tool, with code context significantly improving both lexical and runtime metric performance. Despite its strong performance, \tool has limitations including that it may struggle to generalize to unpopular projects, and the comparison with TeCo likely has data leakage (\tool is likely to have seen TeCo's test set in pretraining). We hope that these limitations can be overcome in future work. Overall, we highlight how incorporating domain knowledge, namely the relationship between code and test files, can be used to create more powerful models for automated test generation. 

\noindent\textbf{Data availability:} The model weights for \tool, code and datasets for training and evaluating \tool, results of additional experiments and comparison with TeCo, CodeGen and StarCoder are available at: \url{https://doi.org/10.5281/zenodo.7901830}. 

\section{Acknowledgements}
The authors would like to thank Charles Sutton for his mentorship as part of the Google Collab Ph.D. Fellowship, which also included \$20,000 in cloud credits without which this work would not have been possible. We additionally thank the authors of TeCo for providing us with data and code for our baseline experiments. \change{We also thank Google Cloud research credits program for the additional \$5,000 in cloud credits that allowed us to further train the model.} This work is supported in part by the US National Science Foundation, awards CCF-2129388 and CCF-1910067.

% The next two lines define the bibliography style to be used, and the bibliography file.
\bibliographystyle{IEEEtran}
\bibliography{references}

\end{document}